\documentclass[12pt]{article}
\usepackage{epsfig}
\begin{document}

\begin {center}
{\Large \bf Meson Spectroscopy without Tetraquarks}
\vskip 5mm
{D V Bugg \footnote{email: d.bugg@rl.ac.uk}\\
{\normalsize  \it Queen Mary, University of London, London E1\,4NS,
UK} \\
[3mm]}
\end {center}
\date{\today}

\begin{abstract}
Data on $e^+e^- \to \pi ^+\pi ^- \Upsilon (1S,2S,3S)$ show a large
increase in branching fractions near $\Upsilon (10860)$.
A suggestion of Ali et al. \cite {AliA} is to interpret this as
evidence for a tetraquark, $Y_b(10890) \equiv (bn)(\bar b \bar n)$.
However, it may also be interpreted in terms of $\Upsilon (10860) \to
\bar B B^*$, $\bar B^*B^*$ and $\bar B_sB^*_s$ above the open-$b$
threshold, followed by de-excitation processes such as $\bar BB^* \to
\Upsilon (1S,2S,3S)$.
In the charm sector, a hypothesis open to experimental test is that
$X,Y$ and $Z$ peaks in the mass range 3872 to 3945 MeV may all be due
to regular $^3P_{1}$ and $^3P_2$ $\bar cc$ states (and perhaps $^3P_0$
and $^1P_1$) mixed with meson-meson.

\end{abstract}

\section {Introduction}
There is much speculation about the existence of tetraquarks.
Most papers unfortunately do not distinguish between tetraquarks and
simple meson-meson.
It is prudent, but less exciting, to examine conventional
interpretations in terms of non-exotic $\bar nn$, $\bar ss$, $\bar cc$
or $\bar bb$ states with admixtures of  meson-meson.

Nuclei may be viewed as six quark combinations.
However, nucleon-nucleon phase shifts may be fitted in terms of
meson exchanges.
The essential reason nuclear matter does not collapse is short-range
repulsion due to the Pauli principle.
Returning to meson spectroscopy, the conventional $\bar qq$ states
appear as octets and singlets (or linear combinations).
Jaffe has suggested that two coloured quarks can form a colour
$\bar 3_c$ which is anti-symmetric, or $6_c$ which is symmetric
\cite {Jaffe}.
Then $0^+$ mesons can be formed as $(\bar 3 3)$ combinations of
flavour and colour.
This neatly evades repulsion between $qq$ and also accounts for the
high masses of $a_0(980)$ and $f_0(980)$. 
However, Jaffe's scheme without meson-meson disagrees with observed 
branching ratios \cite {BRs}.
The ratio $g^2(f_0(980) \to KK)/g^2(a_0(980) \to KK)$ (where $g$ are
coupling constants) requires that $f_0(980)$ has a dominant
$K\bar K$ component; also the predicted ratio 
$g^2(\kappa \to K\pi )/g^2(\sigma \to \pi \pi )$ is much too small.
What may defeat Jaffe's proposal is chiral  symmetry breaking.
The pion and kaon are abnormally light, with the result that
meson-meson configurations $(\bar qq)(\bar qq)$ dominate over
$(qq)(\bar q\bar q)$ in $\sigma$, $\kappa$, $a_0(980)$ and $f_0(980)$.
Present measurements of branching ratios are not  good enough to rule
out some small admixture of $(qq)(\bar q \bar q)$;
further improvements in the branching ratios of $\sigma$ and $f_0(980)$
to $KK$, $\eta \eta$ and $\pi \pi$ above 1 GeV would help greatly.

Meson exchanges do make good predictions for $\pi \pi$ \cite {Caprini}
and $K\pi$ \cite {Descotes} phase shifts up to masses where $q\bar q$
resonances appear.
They also predict correctly low energy $I=2$ and $I=3/2$ phase shifts.
There is no evidence for the $\underline {\bf {27}}$
representation predicted for $(6,6)$ combinations of flavour and
colour.
In a valuable review, Richard \cite {Richard} discusses the
issues in terms of the flux-tubes within four-quark configurations.

There is a further feature relevant to meson-meson configurations.
The train of argument is as follows.
Many mesons are observed at or very close to thresholds of opening
channels.
Well known examples are $f_0(980)$ and $a_0(980)$ at the $KK$
threshold, $f_2(1565)$ ar the $\omega \omega$ threshold and
$X(3872)$ at the $\bar D_0\bar D^*$ threshold within $\sim 0.3$ MeV.
The mechanism for this synchronisation is rather fundamental \cite
{locking}.
The conventional form for the denominator of a resonance is
\begin {equation}
D(s)  = M^2 - s - i\sum _i G^2_i \rho_i(s),
\end {equation}
where $s$ is Lorentz invariant mass squared, $G_i= g_iF_i(s)$, $g_i$
are coupling constants of open channels and $F_i$ are form factors.
However, the correct form for $D(s)$ is
$M^2 - s - {\rm Re} \, \Pi(s) - i {\rm Im}\, \Pi(s)$ where
${\rm Im} \, \Pi(s) = G^2_i(s)\rho_i(s)$ and
\begin {equation} {\rm Re} \, \Pi (s) = \frac {1}{\pi} \rm {P} \int
_{s_{thr}} ^\infty ds' \, \sum _i \frac {G^2_i(s')\rho _i(s') } {s' -
s}.
\end {equation}
Here $s_{thr}$ is the value of $s$ at the opening of the threshold and
P denotes the principal value integral; [${\rm Im} \, \Pi(s)$ is the
pole term from this expression.] 
The origin of Eq. (2) is that amplitudes are analytic functions of 
$s$, so that any change in the imaginary part of the amplitude must be 
accompanied by a change in the real part, or vice versa. 

Fig. 1 sketches the behaviour of ${\rm Im}\, \Pi(s)$ and ${\rm Re}\,
\Pi(s)$ at an S-wave threshold, $KK$ in this example.
The value of ${\rm Re}\, \Pi (s)$  is large and peaks exactly at the
$S$-wave threshold.
It acts as an attractor and can explain why $f_0(980)$ and $a_0(980)$
lie very close to the $KK$ threshold.
The present limitation in calculating $\rm {Re}\, \Pi(s)$
accurately is that form factors $F_i(s)$ are poorly known.
Incidentally, Eq. (2) is equivalent to solving Schr\" odinger type
equations, a procedure adopted by many authors.
These solutions are explicitly analytic. 
It is also included in the model of van Beveren and Rupp, because 
their amplitudes are constructed algebraically to be analytic
\cite {Rupp}.
\begin{figure}[htb]
\begin{center}
\vskip -12mm
\epsfig{file=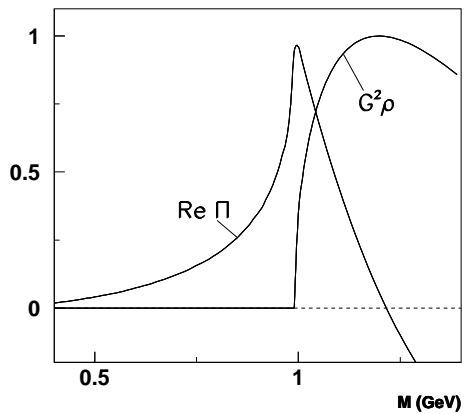,width=8cm}
\vskip -6mm
\caption {${\rm Re} \,  \Pi_{KK}(s)$ and $g^2_{KK}\rho _{KK}(s)$
for $f_0(980)$, normalised to 1 at the peak of $g^2_{KK}\rho _{KK}$.}
\end{center}
\end{figure}

The Hamiltonian for a $\bar qq$ state decaying to meson-meson obeys
\begin {equation}
H \Psi  = \left(
\begin {array} {cc}
H_{11} & V \\
V & H_{22}
\end {array}
\right) \Psi;
\end{equation}
$H_{11}$ describes short-range $\bar qq$ components
and $H_{22}$ refers to ingoing and outgoing mesonic channels and
must include $t$- and $u$-channel meson exchanges; $V$
accounts for the coupling between them due to $s$-channel decays.
The eigenfunction $\Psi$ becomes a linear combination of $\bar qq$
and meson-meson.
The latter is not an `optional extra'; if a resonance decays to
meson-meson, that component is a necessary part of the wave function,
and indeed plays a vital role.
This is the basis of a large number of papers by Oset and
collaborators, enumerating the attractive meson-meson contributions
to a large number of mesons, e.g. \cite {Gamermann}; their calculations include
empirical short-range terms which may well simulate $\bar qq$
contributions.

The form of Eq. (3) is strictly analogous to the formation of covalent
bonds in chemistry \cite {covalent}.
According to the variational principle, the eigenstate minimises the
eigenvalue.
The $\bar qq$ component is of short range.
Mixing with meson-meson components at longer range lowers momentum
components in the wave function and hence the energy eigenvalue.

The $J/\Psi$, $\Psi '(1S)$, $\Upsilon (1S)$ and $\Upsilon (2S)$ are
very narrow and it is a good approximation to view them as pure $\bar
cc$ states.
Some commentators then argue that other mesons cannot be $\bar cc$ or
$\bar bb$ because they have unexpected meson-meson components.
That is wrong.
If a resonance is close to an opening threshold with the same
quantum numbers, it must contain virtual components of that
channel.
Taking $c\bar c n\bar n$ as an example, there are
attractive long range mesonic components in the configurations
$(c\bar n)(\bar c n)$ and $(c \bar c)(n\bar n)$; examples are
$D\bar D^*$ and $\omega J/\Psi $, which appear prominently in
observed decays of X,Y,Z in the mass range 3872 to 3945 MeV.
The observed meson-meson decays are in fact a signature of mesonic
components in the wave function.
Weinstein and Isgur modelled the $f_0(980)$ and $a_0(980)$ in terms of
meson exchanges \cite {Isgur}, as did the Julich group of
Janssen et al. \cite {Janssen}.

These general remarks set the scene for further details.
Section 2 discusses $\Upsilon (10860)$ in this light.
It might be the first clear tetraquark.
However, it may also be understood at least qualitatively as a
$\bar bb$ state coupled strongly to $\bar B B^*$, $\bar B^* B^*$ and
$\bar B^*_s B^*_s$ channels which open in the mass range 10559 to 10826
MeV and can de-excite to $\pi \pi \Upsilon (nS)$, $n<5$.

Section 3 discusses states popularly known as $X,Y,Z$.
The Particle Data Group \cite {PDG} labels them as
$\chi_{c2}(2P)$, $X(3940)$ and $X(3945)$; the last of these is somewhat
confusing in view of the fact that its average mass is now
$3915$ MeV.
Here $Z(3930)$ will be used as a shorthand for $\chi_{c2}(2P)$.
All these states lie close to the $\bar D D^*$ threshold.
This leads to my suggestion that all of these states are
$n=2$ $\bar cc$ $P$-states mixed with meson-meson.

Spin-parity determinations of these states are presently lacking.
This appears to be due to the fact that only decays have been
analysed; that leads to ambiguities.
It is essential to analyse the full process of production and
decay, as in the analysis of Dalitz plots.
This leads to orthogonal matrix elements between all $J^P$.
The ideal process for analysis is $B \to K(\omega J/\Psi )$; here
$B$ and $K$ both have spin 0, making the formulae rather simple.
Formulae for this process were published in 2005 \cite {PWA},
but experimental groups have not taken advantage of spin information
from decays of $\omega$ and $J/\Psi$; that is essential because 
it determines spin alignments.
With the full formulae, my simulation (which assumes uniform geometric
acceptance) suggests that $\sim 50$ events are sufficient to determine
$J^P$ if only a single amplitude is present or dominant.
If there are more amplitudes, it gets harder, but it would still be
valuable to know the outcome.

Section 4 reviews the Partial Wave Analysis using tensor notation in
its simplest form.
The formulae can be used in fully relativistic form.
However, for the mass range 3872 to 3945 MeV, relativistic corrections
are small and barely significant with present statistics.
Two appendices discuss the tensor notation, also the simplest choice of
axes and questions of how to do Lorentz transformations and rotations
of axes, if needed.
Section 6 summarises conclusions briefly.

\begin{figure}[htb]
\begin{center} \vskip -52mm
\epsfig{file=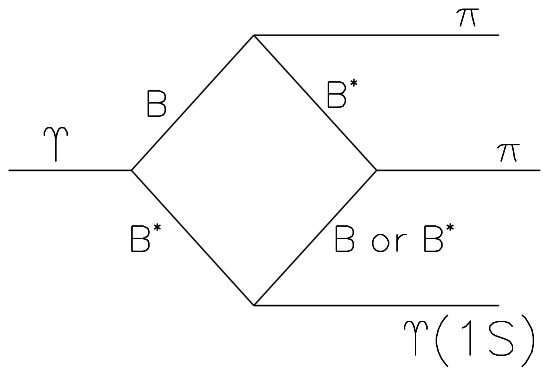,width=15cm}
\vskip -52mm
\caption {One diagram for $\Upsilon (10860) \to \pi \pi \Upsilon (1S)$.}
\end{center}
\end{figure}

\section {The $\Upsilon (10860)$}
The Belle Collaboration reports branching fractions for
$\Upsilon (10860) \to \pi ^+\pi ^- \Upsilon (1S)$,
$\pi ^+\pi^-\Upsilon (2S)$ and $\pi ^+\pi ^-\Upsilon (3S)$ a
factor $\sim  50$ larger than for $\Upsilon (4S)$ \cite {Chen} \cite {Adachi}.
Ali, Hambrook, Ahmed and Aslam interpret these branching fractions in
terms of a tetraquark $Y_b(10890) \equiv (bn)(\bar b \bar n)$
\cite {AliA}.
More recently, Ali, Hambrock and Mishima \cite {AliB} predict for this
configuration an intensity ratio
\begin {equation}
\sigma [\Upsilon (1S)K^+K^-]/\sigma [\Upsilon (1S)K^0\bar K^0] = 1/4.
\end {equation}
For a $\bar b b$ or $\bar B B^*$ composition, this ratio
should be 1, allowing a clear experimental test.

There is a natural explanation of the sizable branching fraction for
$\Upsilon (10860) \to \pi \pi \Upsilon (1S, 2S, 3S)$ in terms of
diagrams like that in Fig. 2.
The intermediate states are off-shell, allowing $\pi$ production, even
though $B^*$ and $B^*_s$ are below the kinematic threshold for pionic
decays.
The final step allows de-excitation of $B^*\bar B$ to lower lying
$\pi \pi \Upsilon $ channels with larger phase space.
Many similar diagrams exist with intermediate $B$, $B^*$, $B_s$ and
$B^*_s$.
Via mixing with meson-meson configurations, the wave function of
$\Upsilon (10860)$ becomes
\begin {equation}
|\Upsilon (10860)> = \alpha |b\bar b> + \sum_{ij} \beta _{ij}|\bar B_iB_j>
+\sum _{n=1,3}\gamma_n |\pi \pi \Upsilon (nS)>.
\end {equation}
However, it is not possible to calculate these branching fractions
accurately for lack of experimental information.
The decay width of $\Upsilon (4S)$ to $\pi \pi \Upsilon (1S)$ is
very weak, probably because it decays almost entirely to
$B\bar B$. 
Transitions between $B$ and $B\pi$ are forbidden by angular momentum 
conservation. 
Channels involving real $B^*$ and $B^*_s$ are closed, so only virtual
intermediate states contribute.
However, once intermediate $\bar B B^*$, $\bar B^* B^*$, $\bar B_s B_s$
and $\bar B_s B^*_s$ channels are open, there are 20 alternative diagrams.
There may be complicated interferences amongst them.
The conclusion is that the observed $\pi^+\pi ^-\Upsilon (1S)$ decays
of $\Upsilon (10860)$ are not dramatically out of line with
expectation, bearing in mind the increase in phase space.
What would be decisive is an accurate determination of the branching
ratio of $\Upsilon (10860)$ to $K^+K^-$ and $K^0\bar K^0$.

A technical detail concerns the fit to the $\pi ^+\pi ^-$ angular
distribution.
The $\pi ^+\pi ^-$ amplitude contains a broad pole at $(440-470)$
MeV.
In some sets of data it appears as a peak, e.g. in
$J/\Psi \to \omega \pi ^+\pi ^-$ \cite {BESII}.
In others it appears as in elastic scattering.
In principle, it therefore needs to be fitted by an amplitude
$[\Lambda _1 + \Lambda _2(s - s_A)]/D(s)$, where $s_A$ is the Adler
zero, 0.0097 GeV$^2$ and $\Lambda$ are coupling constants.
An algebraic parametrisation of  the denominator $D(s)$ is given in
Ref. \cite {sigpole}.
It is not correct to fit the data with the E791 parametrisation, a
Breit-Wigner amplitude of constant width.
That has a phase variation in complete disagreement with $\pi \pi$
elastic scattering and violates Watson's theorem \cite {Watson}.
Furthermore, it contains an unphysical pole just below the $\pi \pi$
threshold.

\section {X,Y,Z}
It is well known to theorists that the opening of strong thresholds
can move resonances by large amounts.
As one example, Barnes \cite {Barnes} has studied charmonium loop
diagrams and concluded that mass shifts are potentially $>100$ MeV; 
he finds that they tend to be similar for all states in a multiplet 
with similar radial excitation number $n$ but different orbital 
angular momentum $L$, e.g. the $n=2$ $^3P_{0,1,2}$ and $^1P_1$ 
states.

The $Z(3930)$ is identified clearly in Belle data for
$e^+e^- \to e^+e^-  + \bar DD$ by a strong $D$-wave component in $\bar DD$
with mass $M=3929 \pm 5 \pm 2$ MeV, $\Gamma = 29 \pm 10 \pm 2$ MeV.
\cite {Z3930}.
It is now adopted by the PDG as the $\chi_{c2}(2P)$ radial excitation
of $\chi_{c2}$ \cite {PDG}.
There are two observations still labelled $X(3945)$ by the
PDG.
One is by Babar in $B \to K(\omega J/\Psi )$ with
$M = 3914.6 ^{+3.8}_{-3.4} \pm 2.0$ MeV,
$\Gamma = 34 ^{+12}_{-8} \pm 5$ MeV \cite {Babar3915}.
The second, with less events $(49 \pm 15)$, is by Belle in
$e^+e^- \to e^+ e^-(\omega J/\Psi )$ with $M = 3916 \pm 3 \pm 2$
MeV, $\Gamma = 17 \pm 10 \pm 3$ MeV \cite {Belle3915}.
The mass difference from $ Z(3930)$ is $14 \pm 6 \pm 4$ MeV, so it is
quite possible this is the $\chi_{c2}(2P)$ decaying to $\omega J/\Psi$.
Another possibility is the $\chi_{c0}(2P)$.

There is considerable scatter in masses predicted for
these states.
A representative collection of recent predictions is shown in Table 1.
Mass differences between $^3P_2$ and $^3P_0$ radial excitations
are large, but can be affected by thresholds and form factors.
If the $n=2$ $^3P_2$ predictions can be trusted, at least
approximately, this state has been pulled down in mass by a large
amount.
The strong signal for $\omega J/\Psi$ at 3915 MeV is a hint of
mixing between $c\bar c$ and the $\omega J/\Psi $ channel
and attraction to this threshold and/or the $D\bar D^*$ threshold.
\begin{table}[htb]
\begin {center}
\begin{tabular}{cccc}
\hline
State & BGS [25] & EFG [26] & [27]
\\\hline
$^3P_2$ & 3972 & 3972 & 3941 \\
$^3P_1$ & 3925 & 3929 & 3900 \\
$^3P_0$ & 3852 & 3854 & 3839 \\
$^1P_1$ & 3934 & 3945 & 3909 \\
$^1D_2$ & 3799 & 3811 & 3799 \\
\hline
\end{tabular}
\caption{Recent predictions of masses of $n=2$ $^3P$ states and
$n=1$ $^1D_2$ in MeV.}
\end {center}
\end{table}

The $X(3945)$ claimed by Belle \cite {Choi} at $3943 \pm 11 \pm 13$ 
MeV with $\Gamma = 87 \pm 22 \pm 26$ MeV does not agree well with 
masses near 3915 MeV.
However, they are in the same decay channel and with a small 
stretch of the errors may both be consistent with $Z(3930)$.

A puzzle is where the $^3P_0$ state is.
Its obvious decay mode is to $\bar DD$.
Its width is strongly dependent on the form factor for the
$\bar DD$ channel \cite {Eichten},\cite {Swanson}.
An early paper of Babar comments that separation of $D\bar D$
events is obscured by mis-identified $\bar D D^*$ events, so it
could have escaped detection just below that threshold 
\cite {Babar03}.
The other obvious possibility is that it could be the 3915 MeV
peak in $\omega J/\Psi$, but then the inference is that it couples
strongly to that threshold.

The $X(3940)$ is observed by Belle in $e^+e^- \to J/\Psi$ + an
inclusive collection of $\bar DD$ and $\bar D D^*$ events.
From 52 identified $\bar D D^*$ events, Belle find a mass of
$M = 3943 ^{+7}_{-6}  \pm 6$ MeV \cite {Pakhlov}.
There are two obvious possibilities for its spin-parity.
Firstly, it could be the missing $^1P_1$ state. 
Secondly, it could be $\chi_{c2}(2P)$ decaying to $\bar D
D^*$ with orbital angular momentum $\ell = 2$, though there is
presently no claim to observe it in $\bar DD$.
The rising phase space for $\ell = 2$ would naturally move the peak
up in mass.
To illustrate this, suppose $Z(3930)$ is described by a Breit-Wigner
resonance of  constant width with Belle parameters.
A simple calculation using $\bar D D^*$ phase space and an
$\ell = 2$ centrifugal barrier for decay with a radius of interaction
of 0.73 fm \cite {FF} gives a peak at 3939 MeV with width 43 MeV;
this is close to Belle parameters, but perhaps fortuitous in view
of experimental errors for both $Z(3930)$ and $X(3940)$.
Also strong coupling to the $\bar D D^*$ channel would move the
peak by an amount which cannot be calculated without knowledge of
the relative branching ratios to $\bar DD$ and $\bar D D^*$.
A puzzle is that Belle do not observe any significant
$B \to K(\bar D D^*)$ signal near 3940 MeV \cite {Aushev}.

The spin-parity analysis of $e^+ e^- \to J/\Psi + 2^+$ involves 
spins $1^- \to 1^- + 2^+ + L$;
since momenta in the final state are large, $L$ can run up to 4.
This makes a complete partial wave analysis impossible with present
data and $J^P = 0^+$ can be confused in this reaction with the 
helicity 0 component of the $^3P_2$ state.

\section {Spin-Parity Analysis}
Spin-parity determinations of these states are needed if any progress
is to be made in understanding them.
The cleanest route to $J^P$ determination lies in
$B \to K + X$, $X \to \omega J/\Psi $ (or $\rho J/\Psi$)  or
$\bar D D^*$.
This is because both $B$ and $K$ are spinless, with the result that
the kaon is produced with known orbital angular momentum equal to
the spin of X; this simplifies the partial wave analysis greatly.
However, that analysis {\it must} use formulae which
describe both production and decay of X and must use spin
information from both $\omega $ and $J/\Psi$ or from the $D^*$.
The spin of the $\omega$ is normal to its decay plane in the rest
frame of $\pi ^+\pi ^-\pi ^0$.
The $J/\psi$ is detected via decays to lepton-lepton, and its
polarisation vector $\epsilon$ is normal to the lepton-lepton axis
in its rest frame.
These are key pieces of information.
With them, matrix elements for the full amplitude contain correlations
between the spectator kaon and decays of $\omega $ and $J/\Psi$ which
are highly distinctive; without these correlations, most of the
relevant information is discarded.
Formulae for these matrix elements were published in 2005
\cite {PWA} and will be reviewed here in outline,
together with comments on the simplest choice of reference frames.

If formulae are written in terms of vectors and tensors, the
vital features are transparent.
The formulae can be written in any reference frame, but in practice
they simplify greatly in the rest frame of X,Y,Z for reasons explained
algebraically in Appendix 1.
A general point is that all matrix elements are orthogonal;
that is a valuable check on computer programmes.

\begin{table}[htb]
\begin {center}
\begin{tabular}{ccc}
\hline
Mass (MeV) & $\bar D D^*$ barrier & $\omega J/\Psi$ barrier\\
\hline
3900 & 0.1905 & 0.1105 \\
3915 & 0.2880 & 0.1835 \\
3930 & 0.3679 & 0.2498 \\
3945 & 0.4361 & 0.3096 \\
3960 & 0.4937 & 0.3635 \\
\hline
\end{tabular}
\caption{$L=2$ centrifugal barriers for $\bar D D^*$ and $\omega J/\Psi$.}
\end {center}
\end{table}
The decay of the resonance $X$ is described by combined spins $s$ of
$\omega$ and $J/\Psi$ (or $D + D^*$).
The $Z(3930)$ is likely to decay dominantly with $\ell = 0$,
though $l=2$ is allowed in principle.
However, the centrifugal barrier factor for $\ell = 2$ is quite
strong.
Table 2 exhibits Blatt-Weisskopf centrifugal barrier factors for the
amplitude using a generous radius of interaction 0.73 fm, including
the convolution of both particles in the final state; (average masses
are used for $D$ and $D^*$).
For a smaller radius, the barrier factors are even smaller.
The $\ell = 4$ barrier makes that amplitude negligible.
The primary objective of the analysis should be to determine the
dominant amplitudes, hence should start with $J^P = 0^+$ and $2^+$,
both with $\ell = 0$.
Further $\ell = 2$ amplitudes may then be introduced one by one with 
penalty functions which constrain their magnitudes unless the data 
really demand them.
This may be done, for example, by adjusting the penalty function
to contribute $\chi^2$ or log likelihood in the range 1--4.

\begin{figure}[htb]
\begin{center} \vskip -35mm
\epsfig{file=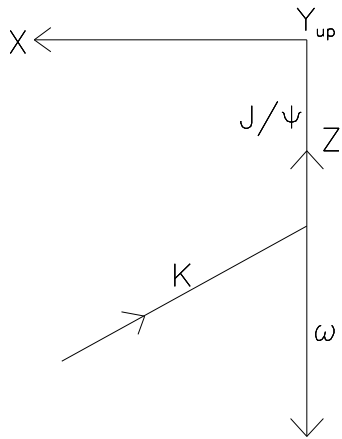,width=10cm}
\vskip -6mm
\caption {Axes describing $B \to K\omega J/\Psi$.}
\end{center}
\end{figure}
A simple choice of axes is shown in Fig. 3 with the $Z$-axis along the
direction of the $J/\Psi$ and $Y$ along the normal to the plane of
K, $\omega $ and $J/\Psi$.
The formula for $\omega \to \pi ^+\pi ^-\pi^0$ is a 4-vector $W$
given in Lorentz invariant form in Section 4.2 of \cite {PWA}.
This formula is easily evaluated in any reference frame, such as that
of Fig. 3.
It includes naturally the well-known enhancement near the edge
of the Dalitz plot for its decay to $\pi ^+\pi ^-\pi ^0$.

The $J/\Psi$ decay is simplest in its rest frame.
The effect of the Lorentz transformation to the rest frame of
X,Y,Z is small because the $J/\Psi$ is heavy and therefore
non-relativistic in that frame.
The gymnastics required for the Lorentz transformation of its
polarisation vector $\epsilon$ are given in Appendix 2.

For $J^P = 1^+$, the kaon in the production process carries $L = 1$.
In the resonance rest frame, this angular momentum is described by
the 3-momentum $\vec {K}$ of the kaon.
In the decay $X(3872) \to \rho J/\Psi$, the spin 1 of the $\rho$
is represented by the vector $P = k(\pi _1) - k(\pi _2)$,
where $k$ are momenta in any reference frame.
In the non-relativistic limit the full matrix element reduces to
$\sl {M} = \vec {\epsilon}.\vec {P} \wedge \vec {K}$.
There is a simple trick which can be used to do the spin average
over $\vec {\epsilon}$.
This is to write $\epsilon _x = 1$ and $\epsilon _y = i = \sqrt {-1}$;
then intensities are obtained by taking the modulus squared of
$\sl {M}$.
For $X(3872)$, $\ell = 2$ in the decay is eliminated by the
tiny width of the state.
The threefold dependence on the lepton axis, the spin of the $\rho$
and the kaon direction is then unique and highly distinctive.

For $J^P = 0^+$, the matrix element is
\begin {equation}
M  = \vec {W}.\vec {\epsilon} = W_1\epsilon _1 + W_2\epsilon _2 +
W_3\epsilon _2 - W_4 \epsilon_4,
\end {equation}
where $\vec {W}$ and $\vec {\epsilon}$ are 4-vectors.
Without the last term, which is very small, the matrix element is
just the scalar product of 3-vectors $\vec {W}$ and 
$\vec {\epsilon}$.
With this simplification, the intensity is $\sin ^2 \alpha$, where
$\alpha$ is the angle between $\vec {W}$ and the lepton axis.
The simplicity of this formula may actually make the $0^+$ state
harder to identify than other amplitudes.

For $J^P = 2^+$, the kaon carries $L = 2$ and is described by a
tensor
\begin {equation}
\tau _{\alpha \beta} = K_\alpha K_\beta - (1/3)\delta_{\alpha \beta }
(K^2).
\end {equation}
Here $\delta$ is the usual Kronecker delta function and
$K^2 = K_1^2 + K^2_2 + K^2_3 - K^2_4$.
For decays with $\ell = 0$, the combined spin of $\omega$ and
$J/\Psi$ is $s = 2$.
The decay is then described by the tensor
\begin {equation}
T_{\alpha \beta} = \epsilon _{\alpha} W_\beta + \epsilon _\beta W_\alpha
 - (2/3)(\epsilon_\mu W^\mu )
\end {equation}
in the resonance rest frame;
$\epsilon _\mu W^\mu$ implies the summation over $\mu = 1$ to 4 and
takes the same general form as $K^2$ above.
The full matrix element is just the contraction
$\tau _{\alpha \beta }T^{\alpha \beta }$.
It is very distinctive because of the dependence on angles of all of
$K$, $\vec {W}$ and $\vec {\epsilon}$.

A simulation of these $J^P$, assuming full geometrical acceptance,
suggests that 50 events would give a separation at $90\%$ confidence
level between $0^+$, $1^+$ and $2^+$ if only one of them
contributes a significant amplitude.
Babar and Belle between them have many more events than this, with the
exception of $X(3940) \to \bar D D^*$.
If more than one $J^P$ contributes, the separation is obviously poorer,
and an overlap between $0^+$ and $2^+$ could be a difficult case.
Formulae for $\ell = 2$ in decays are given in Ref. \cite {PWA}.

A distinctive feature of $X(3872)$ is that its mass lies within 0.3 MeV
of the $\bar D_0 D^*_0$ threshold.
There is a simple explanation.
It is attracted to this threshold by the dispersive effect
due to the opening of the threshold.
Lee et al. \cite {Lee} model the $X(3872)$ including exchange of $\pi$
and $\sigma$ between $\bar D$ and $D^*$.
They solve the Schr\" odinger equation, including the $\bar D D^*$
channel; this is equivalent to evaluating the dispersion integral
of Eq. (2), since their solution is explicitly analytic. 
One should not be misled into thinking that the effect of
$\bar D D^*$ is small because few events are seen in 
$X(3872) \to \bar D_0 D^*_0$; there are
few events because of the narrow width of $X(3872)$.
However, the attraction to the threshold does not involve
the resonance denominator in Eq. (2); it arises from the integral over 
all phase space for $\bar D D^*$.
A parallel example is the deuteron.
Its binding energy is not governed by its tiny width; it comes from
virtual meson exchanges over the whole left-hand cut.

Present data have failed to distinguish between $J^P = 1^+$ and $2^-$
for $X(3872)$.
The key point is to check whether partial wave analysis is consistent
with $J^P = 1^+$ or not.
If it is, the fact that the mass coincides with the $\bar D D^*$
threshold is significant information.
A state with $J^P = 2^-$ couples to $\bar D D^*$ with $l=1$ and there
is no obvious reason why it should lie close to this threshold.
In fact, predictions for the mass of $\eta _{c2}(1\, ^1D_2)$ in Table 1
lie far below $X(3872)$; this is because spin-splitting for $D$ states
is smaller than for $P$-states and the mass of the $\psi (3770)$
(mostly $n=1\, ^3D_1$) is well known.
For $J^P=2^-$, formulae are more messy and are given in Ref.
\cite {PWA}.
There are are in principle three amplitudes, but one requires $\ell = 3$,
which is irrelevant for the narrow $X(3872)$.
The others have $\ell = 1$ and combined spins either $s = 1$ or 2.

A full partial wave analysis of Belle data for $X(3940) \to \bar D D^*$
will in general be ambiguous with only $\sim 50 $ events.
However, this may be enough events to test the possibility that
$J^P = 0^-$.
The spin information is carried by the vector
\begin {equation}
Q = k_D -  k_\pi - \frac {M^2_D - m^2_\pi }{M^2_{D^*}}[k_D + k_\pi],
\end {equation}
where $k$ are momenta (in any frame).
With $J^P=0^-$, production requires $L=1$ and decays to $D\bar D^*$
with $\ell = 1$, so only a single amplitude is present; it has the
distinctive feature of going through zero at the centre of the
angular distribution for decays.
There is a missing $^1P_1$ $\bar cc$ state. 
From predicted masses in Table 1 and the fact that the mass of
the $n=1$ $^1P_1$ state is slightly about the $^3P_1$ state, it seems
likely that the $^1P_1$ state will lie above $X(3872)$.
However, it is not easy to find. 
It has charge conjugation $C=-1$, while $\omega J/\Psi$ has $C=+1$,
so this decay is forbidden. 
It can appear in $\bar D D^*$.
However, it appears in that channel in the combination
$\bar D_+ D^*_- + \bar D_0 D^*_0$ (because of the minus sign associated
with $\bar D_0$), while $^3P_1$ decays with the opposite relative 
sign. 
Without interferences with a state of established $J^{PG}$, $^1P_1$
and $^3P_1$ are not separated in any particular charge state. 

\section {Conclusions}
It is necessary to be specific about what tetraquark configuration
is meant in every discussion of them.
Meson-meson combinations mixed with $q\bar q$ are to be expected via
$t$- and $u$-channel exchanges and also from the $s$-channel decay.

The increase in branching fractions for
$e^+e^- \to \pi ^+\pi ^- \Upsilon (1S,2S,3S)$ as the mass increases
from $\Upsilon (4S)$ to $\Upsilon (10860)$ may be understood
qualitatively via the opening of real decays to $\bar B B^*$,
$\bar B^* B^*$ and $\bar B_s B^*_s$ channels in the mass range
10559 to 10826 MeV, followed by de-excitation to $\pi\pi\Upsilon$.
Observed branching fractions are not seriously out of line with
expectation.
The composition $(bn)(\bar b\bar n)$ may be tested by measuring the
intensity ratio given by
$\sigma [\Upsilon (1S)K^+K^-]/\sigma [\Upsilon (1S)K^0\bar K^0)$,
which Ali, Hambrock and Mishima show to be 1/4 for exotic tetraquarks;
for a $\bar B B^*$ composition or $\bar bb$, this ratio should be 1.

Spin-parity analyses of X,Y,Z states are needed and can be done
straightforwardly in several cases using formulae developed in Ref.
\cite {PWA}.
It is suggested here that they can be explained naturally as the
expected $n=2$ radial excitations of $^3P_1$, $^3P_2$ and perhaps
$^3P_0$.
It is essential to make use of the full forms of matrix elements for
production and decay for partial wave analysis.

\section {Acknowledgement}
I am grateful to Prof. George Rupp for valuable comments on
the points discussed here.
\section {Appendix 1}
Chung gives an introduction to tensor formalism \cite {Chung}.
Foundation work was done by Zemach \cite {Zemach}.
The convention used here is that the Lorentz metric $g_{\mu \nu}$ is
such that four-vectors are written as $(-p_1, -p_2, -p_3, E = p_4)$,
i.e. $g_{\mu \nu}$ has only diagonal elements (-1,-1,-1,1).

The decay of a scalar particle, e.g. $\sigma \to \pi \pi$, is described
by its mass $(k_1+k_2)_4 \equiv P_4$; here $k_{1,2}$ are 4-vectors of
the decay pions.
A spin 1 particle needs to be described by a vector constructed to be
orthogonal to the scalar expression.
This is done, for example in describing $\rho \to \pi \pi$, using
\begin {equation}
(k^\bot _{12})_\mu = g^\bot _{\mu \nu }(k_1 + k_2)_\nu,
\end {equation}
where
\begin {equation}
g^\bot _{\mu \nu} = g_{\mu \nu} - (P_\mu P_\nu)/M^2_\rho.
\end {equation}
Substituting (11) into (10),
\begin {equation}
(k^\bot _{12})_\mu = g_{\mu \nu}(k_1 - k_2)_\nu -
\frac {[(k_1-k_2)_\nu (k_1 + k_2)_\nu](k_1 + k_2)_\mu}{M^2_\rho}.
\end {equation}
The orthogonality between vector and scalar is demonstrated from the
dot product of $(k^\bot _{12})_\mu$ with $(k_1 + k_2)_\mu$:
\begin {equation}
g_{\mu \nu}(k_1-k_2)_\nu(k_1+k_2)_\mu - (M_1^2 -
M_2^2)M^2_\rho/M^2_\rho = 0.
\end {equation}

In the rest frame of the decaying particle, $g^\bot _{\mu \nu}$ has
diagonal elements [-1,-1,-1,1]-[0,0,0,1]=
\newline
-[1,1,1,0].
This gives the important result that $k^\bot$ is a 3-vector in the
rest frame of a resonance.
Then $k^\bot$ describes the angular momentum carried by the particle.
Tensor expressions for higher spins are built up as described by
Zemach and Chung so that they are orthogonal to lower spins.
They are symmetric and traceless, i.e. diagonal elements add to 0.
We shall need only the tensor $T_{\mu \nu}$ for spin 2.

\section {Appendix 2}
Suppose the lepton pair from $J/\Psi$ decay has components
$(\sin \theta \cos \phi$, $\sin \theta \sin \phi$, $\cos \theta )$
in axes in the $J/\Psi$ rest frame parallel to $X,Y,Z$ of Fig. 3.
This vector may be expressed as
\begin {equation}
\left(\begin {array} {ccc}
\cos \phi & -\sin \phi & 0 \\
\sin \phi &  \cos \phi & 0 \\
0 & 0 & 1 \\
\end {array}
\right)
\left(\begin {array} {ccc}
\cos \theta & 0 & \sin \theta \\
0 & 0 & 1 \\
-\sin \theta & 0 & \cos \theta \\
\end {array}
\right)
\left(
\begin {array} {c}
0 \\
0 \\
1\\
\end {array}
\right).
\end {equation}
The spin $\epsilon$ of the $J/\Psi$ is orthogonal to this lepton axis.
In axes x,y,z aligned along the lepton axis, $\epsilon$ is given by
a vector $(\cos R, \sin R, 0)$ and eventually it is necessary to
average over $R$.
In X,Y,X, axes, $\epsilon $ is given by the converse relation to Eq.
(14) with the result
\begin {equation}
\left(\begin {array} {ccc}
\cos \theta & 0 & -\sin \theta \\
0 & 1 & 0 \\
\sin \theta & 0 & \cos \theta \\
\end {array}
\right)
\left(\begin {array} {ccc}
\cos \phi & \sin \phi & 0 \\
-\sin \phi &  \cos \phi & 0 \\
0 & 0 & 1 \\
\end {array}
\right)
\left(\begin {array} {c}
\cos R \\
\sin R \\
0   \\
\end {array}
\right)
\end{equation}
\begin {equation}
 = \left(\begin {array} {c}
\cos \theta (\cos \phi \cos R + \sin \phi \sin R) \\
-\sin \phi \cos R + \cos \phi \sin R\\
\sin \theta (\cos \phi \cos R + \sin \phi \sin R)\\
\end {array}
\right).
\end {equation}
Under the Lorentz transformation back to the centre of mass frame,
this becomes
\begin {equation}
\left(\begin {array} {c}
\cos \theta (\cos \phi \cos R + \sin \phi \sin R) \\
-\sin \phi \cos R + \cos \phi \sin R\\
\gamma \sin \theta (\cos \phi \cos R + \sin \phi \sin R)\\
\beta \gamma \sin \theta (\cos \phi \cos R + \sin \phi \sin R) \\
\end {array}
\right),
\end {equation}
where $\gamma$ and $\beta$ are the usual expressions for the Lorentz
transformation.
This approach is simpler than that given in Ref. \cite {PWA}.

\end  {document}